# The future of blockchain-enabled tokenization in infrastructure investment and development: A Delphi-based scenario analysis

**Yifeng Tian**[1], **Chaofeng Wang**[1], **Junghoon Woo**[1], **Zheng Lu**[2], **and Peter Adriaens**[3]

[1] College of Design, Construction, and Planning, University of Florida, Gainesville, FL 32603, USA
[2] Center for Buildings, Infrastructure, and Public Space, Columbia University, New York, NY 10025, USA
[3] Department of Civil and Environmental Engineering, University of Michigan, Ann Arbor, MI 48109, USA

**Abstract**

Spurred by emerging blockchain technology and increased interest in tokenization, this forecasting research is built on extensive literature and aggregated expertise to explore the potential implementation of blockchain-enabled tokenization in infrastructure investment and development. The Delphi-based scenario analysis approach was applied to gather long-term forecasts and assessments of a research panel consisting of 39 experts in blockchain tokenization and infrastructure development on how tokenization will influence the future of infrastructure finance and identify scenarios of potential applications and impact. International experts were segregated into two groups salient to this topical area based on both experience ad self-identification: infrastructure development and blockchain tokenization. Twenty-three projections for 2035, developed from a literature review, case study analysis, and expert interviews, concerning perspectives of both the supply and demand side for the adoption of blockchain tokenization, were assessed in a two-round Delphi analysis. Regulatory, economic, social, and technological perspectives of tokenization were taken into consideration. Assessments were based on both probability and impact of occurrence. Three groups of scenarios resulted from quantitative and qualitative analysis, reflecting agreement and differentiation between both expert groups. The results of this study clearly underlined the potential of tokenization in infrastructure. Uncertainties and barriers confronting the technologies' diffusion were discussed. This study contributes to the transfer of general technical-driven blockchain-enabled tokenization knowledge to infrastructure-specific tokenization knowledge. Long-term strategic planning is supported by this study with the scenario data acting as a starting point for blockchain-related efforts in infrastructure development.

## 1. Introduction

Infrastructure is essential for social development and economic expansion. From power generation facilities to transport systems and water networks, infrastructure provides foundational services which enable society to function and economies to thrive. Infrastructure plays a vital role in strengthening inclusiveness and sustainability (Gupta and Vegelin, 2016). Investment in infrastructure is one of the main drivers to generate long-term growth and stimulate economies out of recession after a systemic crisis such as the COVID-19 pandemic and contributes to the United Nations Sustainable Development Goals (SDGs) (Thacker et al., 2019). Even though infrastructure is at the nexus of economic and social prosperity, there is an increasing mismatch between the need for more infrastructure and available financing globally, further accelerating socio-economic disparities. According to the World Economic Forum's forecast, the world was facing a $15 trillion infrastructure gap by 2040 (Davisson and Losavio, 2020). The World Bank estimated that developing countries needed to triple the current annual spending on infrastructure over the next decade (Mapila et al., 2017). The demand for investment in infrastructure will only increase with time.

Under current public sector financing models for infrastructure, municipal bonds, direct grants, subsidies, and concessionary loans are subject to political influences and budgetary constraints. These challenges are exacerbated by the emergent pandemic spending and tighter financial regulations (e.g., Basel III) that have limited availability and access to traditional public capital sources to fund infrastructure (Estache, 2010; Esty, 2014; Humphreys et al., 2018). Private investments in infrastructure, typically through debt (loans and bonds) and direct (equity) investment, are only available for infrastructure projects and limited to a narrow set of institutional investors such as pension funds and insurance companies. This leaves a significant number of private resources and capital on the sidelines (Yescombe and Farquharson, 2018). Current infrastructure finance models face further challenges and increasing regulatory requirements to integrate environmental, social, and governance (ESG) factors to facilitate long-term sustainable development (Thacker et al., 2019; Inderst, 2020; Gainfrancesco, 2021; Tian et al., 2022a). Therefore, pioneering thinking and innovative financing approaches and models will be required to not only address public sector funding limitations and mobilization of private financial resources but, more importantly, to bridge the widening infrastructure gap and realize inclusive and sustainable growth to positively impact society, the economy, and the environment.

Since the advent of cryptocurrencies in 2009, the applications for blockchain, a combination of distributed ledger technology (DLT) with encryption technologies, have soared (Nakamoto, 2008; Sunyaev, 2020). Blockchain-based systems are designed to maintain and transact tamper-proof and permanent records of data. With the advancements in the blockchain, an immutable, decentralized, and trusted system can be built, bringing improved security and transparency (Swan, 2017). Over the last decade, blockchain has evolved from a promising concept to a technology with applications ranging from corporate supply chain management to decentralized financial markets, billed for the present and future. Blockchain has been causing major stirs in various industries and inspiring the creation of new business models (Morkunas et al., 2019; Woo et al., 2021). Building on the blockchain, tokenization enables the transition of assets with values in conventional forms or access rights into cryptographic tokens (Morrow & Zarrebini, 2019; Khan et al., 2020). The transition could improve efficiencies by orders of magnitude (e.g., cost of finance, transparency, liquidity). Even though the application of tokenization in infrastructure development is scarce at present, it has shown great potential to serve as an alternative financing vehicle to supplement the current infrastructure development and finance system to bridge the gap (Tian et al., 2020).

Despite the growing attention from industries and markets, potential applications of tokenization in infrastructure projects have not attracted comparable academic interest. In part, this is the result of a lack of theoretical and empirical analysis. In light of the technology's apparent gap regarding the future development and high degree of uncertainty, especially in terms of the political, economic, social, and technological aspects of blockchain tokenization for infrastructure investment and development, this study builds upon the extensive research practices within the

Delphi-based scenario planning literature. The emphasis was to contribute to a systematic development of expert opinion consensus in an area that is essential for the realization of tokenization in infrastructure investment and development. In the following sections, relevant literature will be reviewed to discuss the critical factors related to the tokenization of infrastructure. Research methods are explained in Section 3. Subsequently, the research findings are examined, and conclusions are drawn.

## 2. Literature Review

### 2.1. Blockchain-enabled tokenization

Tokenization has been used to describe a process where a sensitive data element is substituted with a non-sensitive equivalent, referred to as a token, which has no extrinsic or exploitable meaning or value (Iwasokun et al., 2018). In this case, tokenization is similar to encryption (Stapleton & Poore, 2011). With the advent of DLT, tokenization refers to the process of converting assets with value or access rights into cryptographic tokens on a blockchain (Nassr, 2020). In this context, tokenization is similar to securitization where assets are pooled to be repackaged into new valuable securities. Cryptographic tokens can represent assets as stores of value, such as corporate equity, project bonds, and funds; or represent access rights, such as permissions to a platform or services. In theory, any assets or rights can be tokenized and represented on a blockchain (Heines et al., 2021). Tokenization builds connections between the off-chain world and the on-chain world, where transaction efficiency, information recording, and sharing are expected to be improved (Wandmacher, 2022).

Cryptographic tokens are governed by smart contracts, which are software algorithms with trigger actions based on predefined parameters coded on the blockchain. Once smart contracts are programmed, they are self-executing and self-enforcing (Zou et al., 2019). The automation enabled by smart contracts reduces the number of intermediaries and reduces administrative burdens, which results in faster execution and cost reduction. Other benefits associated with tokenization include improving liquidity for currently illiquid assets (e.g., infrastructure assets), expediting efficient clearing and settlement, and enhancing transparency (Woo et al., 2020; Wang et al., 2018). Cryptographic tokens can be grouped into three categories: payment, utility, and security tokens. Payment tokens are cryptographic currencies for making and receiving payments on the blockchain, such as Bitcoin and Litecoin, serving as alternative means of exchange in addition to fiat currency. Utility tokens, offered by issuers, grant access rights to products or services to token holders. This type of token is primarily used within the blockchain ecosystem. Security tokens are investment instruments representing legal ownership of financial assets (bond, equity, loan, etc.) or physical assets (commodities, real estate properties, etc.) that have been verified on the blockchain (Lambert et al., 2021).

### 2.2. Infrastructure finance

Public and private finances are essential in infrastructure investment and development. For publicly financed projects, they rely on the budget of governments for financing, which takes the form of direct grants, subsidies, credit enhancements, or low-interest loans (Fay et al., 2021). For projects financed by sub-sovereign government agencies or administrations, municipal financing instruments, such as general obligations, anticipated notes, or certificates of participation are commonly used (Croce et al., 2015). The payment mechanisms to service the outstanding debt are based on taxes, fees (e.g., gas taxes), or user-levied fees (e.g., tolls, usage fees). Debt financing (bonds and loans) forms the largest category of private-sector finance in infrastructure investment. Debt instruments can take the form of direct loans held on the balance sheets of financial institutions (full recourse) or be structured for resale to investors in private (e.g., private-placement bonds) or public markets (e.g., registered corporate bonds). Infrastructure project debts can be tailored to fit specific demands and preferences of investors (limited recourse). Equity investors can invest in infrastructure by purchasing listed equities (indirect participation rights in infrastructure corporations or funds) or unlisted equities (direct ownership, management of project assets or corporate entities) (Tang et al., 2010). Mezzanine finance instruments for infrastructure include subordinated loans and bonds (Dong et al., 2011). Corporate finance is the dominant channel in financing private infrastructure (Shirai, 2004). Investments are against the balance sheets of the project developers. In recent years, the financing of larger infrastructure assets has increasingly taken the form of project finance, to reduce financial liability to the corporate developer. Rather, this risk mitigation strategy is backed by the project company's own balance sheet. This technique has emerged to be a financial solution for infrastructure involving public entities in the role of either counterparty or regulator and attracting private capital in high specificity and intensity of capital projects (Brealey et al., 1996; Yescombe, 2013).

Table 1: Taxonomy of infrastructure financing instruments and vehicles (Tian et al., 2020; modified from Croce et al., 2015)

| Modes | | Infrastructure Finance Instruments | | Market Vehicles |
|---|---|---|---|---|
| Asset Category | Instrument | Infrastructure Project | Corporate Balance Sheet/Other Entities | Capital Pool |
| Fixed Income | Bonds | Project Bonds | Corporate Bonds, Green Bonds | Bond Indices, Bond Funds, ETFs |
| | | Municipal, Sub-sovereign bonds | | |
| | | Green Bonds, Sukuk | Subordinated Bonds | |
| | Loans | Direct/Co-Investment lending to infrastructure project, Syndicated Project Loans | Direct/Co-Investment lending to infrastructure corporate | Debt Funds (GPs) |
| | | | Syndicated Loans, Securitized Loans (ABS), CLOs | Loan Indices, Loan Funds |
| Mixed | Hybrid | Subordinated Loans/Bonds, Mezzanine Finance | Subordinated Bonds, Convertible Bonds, Preferred Stock | Mezzanine Debt Funds (GPs), Hybrid Debt Funds |
| Equity | Listed | YieldCos | Listed Infrastructure & utilities stocks, Closed-end Funds, REITs, IITs, MLPs | Listed Infrastructure Equity Funds, Indices, trusts, ETFs |
| | Unlisted | Direct/Co-Investment in infrastructure project equity, PPP | Direct/Co-Investment in infrastructure corporate equity | Unlisted Infrastructure Funds |

Abbreviations: ABS, asset-backed security; CLOs, collateralized debt obligations; GPs, general partners; REITs, real estate investment trusts; IIT, infrastructure investment trust; MLP, master limited partnership; ETF, exchange-traded fund.

2.3. Tokenization in infrastructure investment and development

The tokenization of infrastructure refers to the process of converting information available on utilities (e.g., services or products offered by infrastructure facilities or project developers) or ownership/economic interests of the securities (e.g., equity of infrastructure companies, funds, loans, or bonds) into digital (cryptographic) tokens on the blockchain. Given the increasing desire to transition infrastructure facilities into intelligent systems and the desire to unlock efficient financing, tokenization may facilitate an alternative financing model that can overcome obstacles that have yet been addressed by conventional financing instruments.

Tokenization has been applied in the sector of energy infrastructure (Mengelkamp et al., 2018 ab; Adriaens and Ajami, 2021; Tian et al., 2022a). Tokens representing the access right to electricity trading platform, developer companies' equity shares, or power-purchase agreements are issued to raise capital for future corporation and projects development. In theory, tokenization could be applied in other sectors. Utility tokens backed by tolls can be issued to raise funds for transportation projects. Bond-like tokens (security tokens backed by project, corporate, or municipal bonds) can be issued to raise funds for most any type of infrastructure project (public and private). Depending on the project, whether it generates revenues, bond-like tokens can be paid back by project revenues or availability payments backed by the government in PPP projects (Tian et al., 2020b). Tokenization financing could be easily incorporated with existing corporate and project finance models. Tokens backed by project equity (in PPP projects) or project debts could be issued on a project level. There have been no projects integrating tokenization with project finance at present.

The implementation of tokenization in infrastructure investment and development might potentially improve investment liquidity, transparency, transaction efficiency, and encourage private participation. Data generated in token activities (e.g., issuance, transfer, and settlement) are immutably stored on the blockchain and made available to governments, project administrators, project sponsors, investors, and surrounding communities in real-time from the perspective of technology (access to certain data might be restricted by regulation and contracts). Real-time access to immutable data simplifies the due diligence process. The disintermediation enabled by smart contracts could result in a significant capital cost reduction, which improves the bankability of projects (Schar, 2021). By tokenizing security or utility into small-value tokens, which represent a small portion of economic benefits, investment in infrastructure with a small amount of money becomes financially viable. The lowered barriers to entry would attract small investors, including retail and small and medium-sized enterprises (SME) investors, who were excluded from the infrastructure investment historically in the conventional financing models. By enabling investing with a small amount of money, surrounding community residents of a project might be offered the opportunity to be actively engaged in project financing. They can serve as special shareholders to influence the planning, financing, and future operations of the project through voting rights (approved by regulators and project developers) embedded with the tokens. Consumerization through tokenization thus would be able to create a sense of ownership in local public infrastructure facilities and provides an option to galvanize social acceptance. New economic models enabling the conversion of non-financial values such as positive social and environmental impacts into security or utility tokens, which can be monetized and traded, are also made impossible with tokenization. Even though tokenization could bring efficiency gains to infrastructure investment and development, it is still in its very easy stage of development. Some efficiency gains might only be realized in theory. Challenges of regulations (e.g., unclear legal statutes, cross-border uncertainties, and lack of single point of accountability) and technology (e.g., network stability, interoperability, and cyber-risks) are preventing tokenization from realizing large-scale adoption. It is important to thoroughly understand the new technology in-depth through an empirical analysis.

## 3. Methodology

The Delphi technique has been used for technological forecasting research and is particularly suitable for long-term research objectives (15-30 years), making it a highly suitable method for future research (Gray and Hovav, 2008; Rowe et al., 1991; Skulmoski et al., 2007;). In this study, the technique is applied to forecast and develop future scenarios regarding blockchain-enabled tokenization's application and impact on infrastructure investment and development. The technique has been applied in academic scholarly work to investigate blockchain-related subjects in future research (Schlecht et al., 2021; Holotiuk et al., 2019; Aldrighetti et al., 2021; Schweizer et al., 2020). While the technique was originally implemented to yield consensus among experts regarding specific propositions, the purpose of this study is to foster well-grounded group opinions

(Landeta, 2006; Schlecht, et al., 2021; Kopyto, et al., 2020; Fritschy and Spinler, 2019). The general design of Delphi builds on four principles: anonymity, iteration, controlled feedback, and statistical group response (Vermon, 2009). Properly conducted Delphi studies show higher accuracy in forecasts than conventional surveys (von der Gracht, 2008; Rowe and Wright, 1999) and have proven their reliability and validity (Parenté and Anderson-Parenté, 2011; Landeta, 2006).

Scenario planning is particularly suited for assessing future developments, decision-making under uncertainty, and long-term planning (Varum and Melo, 2010). Scenarios contribute to practitioners and policymakers executing innovation and effectively dealing with situations of strong agreement (Bood and Postma, 1997; Beiderbeck et al., 2021). Scenarios can also be used as a tool to develop theories that are innovative, less formulaic, and disconfirms assumptions. Many scholars recommend the development of Delphi-based scenarios for long-term oriented and explorative future research. In this context, Delphi delivers reliable data that can be easily integrated into the process of scenario construction (Linz, 2012; Novack et al., 2011; Banuls and Turoff, 2011).

Despite its advantages, the application of the Delphi technique has drawbacks. For example, the reliability and validity of the Delphi study are criticized in some cases (Goodman, 1987). Following Kastein et al. (1993) and Schlecht et al., (2021), standardization is applied throughout the entire study to ensure the reliability of its outcomes by: (1) systematically recruiting experts to establish the research panel, (2) retaining the identical questionnaire in both execution rounds, and (3) querying the same participants from the previous round for their participation in the next round. To ensure validity, particular emphasis was placed on the questionnaire development, and on following a systematic approach based on Gnatzy et al., (2011) and Kopyto et al. (2020).

3.1. Development of Projections

The projection development is essential for ensuring the value, validity, and reliability of Delphi-based studies. Rigorous procedures were performed to develop thought-provoking and concise projections, including: data collection, projection formulation, and pretesting (Fristschy and Spingler, 2019; Spickermann, et al., 2014; Heiko and Darkow, 2010). To streamline the phase of data collection, projections were developed through four diverse data sources: desk research, case studies, interviews, and unstructured brainstorming. Relevant journal and conference articles along with articles and posts available on websites were screened to identify relevant factors concerning the future of tokenization in infrastructure investment and development. Keywords such as "blockchain tokenization", "asset tokenization", "infrastructure investment", and "decentralized finance" are used. Additionally, twenty-one projects related to energy infrastructure tokenization were analyzed to detect further factors and to triangulate the screening process, ensuring that relevant issues were addressed in the development of projections. In addition, five semi-structured interviews were conducted with experts from the field of tokenization or infrastructure to enrich the database.

Following this process, a long list of 254 factors concerning the future of tokenization infrastructure was generated. The long list was then condensed into a short list by two members of the research team through deletion of duplicated factors and aggregation, according to Kopyto et al. (2020). Sixty-nine factors remained after consolidation. An internal workshop for projection development was conducted with four members of the research team to cross-validate a short list of factors (42 factors remained) and to develop preliminary projections. Initially, 32 provisional projections were derived. A projection consolidation process was applied based on the following criteria: an appropriate number of elements in each projection (Salancik et al., 1971), clear descriptions of definition and concepts (Johnson, 1976; Wright et al., 2019), and elimination of conditional statements (Roßmann et al., 2018). Due to a large number of projections would decrease the probability of the completion of the questionnaire and lead to a reduced response rate (Parenté and Anderson-Parenté, 1987), the number of projections was further decreased to 23 while maintaining the breadth and depth of topics relevant to the focus of the study (Jiang et al., 2017).

As proposed by Kopyto et al. (2020), the Delphi questionnaire was pre-tested with five academics with deep methodological and/or subject-specific knowledge to check the projections' precision, plausibility, completeness, and methodological soundness. Their feedback resulted in the revision of the projection questionnaire. The final questionnaire included 23 projections aligned with the political, economic, social, and technological (PEST) framework regarding the future of tokenization in infrastructure for the year 2035. A time horizon of 13 years (until 2035) was chosen to stimulate creative thinking and long-term forecasting and is aligned with compatible market projections for the adoption of new technologies. Validated by Nowack et al. (2011) and Heiko and Darkow (2010), this time horizon has been proven appropriate for this type of research (Bokrantz et al., 2017)

Introductions and explanations of the research design are communicated to the selected panelists (experts). The first part of the questionnaire emphasizes basic socio-demographic information, such as years of experience, and specific expertise in blockchain tokenization or infrastructure development. The second part of the questionnaire focuses on projections categorized based on the classical PEST analysis framework, which has been used frequently in foresight studies (Wilson and Gilligan, 2012; Jiang et al., 2017)

### (a) Political and regulatory projections

Legal and regulatory certainty is of great importance for the future application of tokenization in infrastructure investment and development. However, the regulatory uncertainty at the current stage is one of the most serious impediments restraining the broader adoption of the technology (Schletz et al., 2020). Countries like Sweden and the United Kingdom have been making progress in regulating crypto markets (Garcia-Teruel and Simón-Moreno, 2021), and the US is starting this process. In order to further investigate the potential influences of the political and regulatory requirements and the possibility of regulating tokenization-related activities in infrastructure, five projections included in this section are:

Projection 1. In 2035, legal and regulatory frameworks for the tokenization of infrastructure will have been established in most (developed and developing) countries.

Projection 2. In 2035, an international regulatory framework for the tokenization of infrastructure will have been developed.

Projection 3. In 2035, legal obligations and regulations of smart contracts will have been clarified in most countries.

Projection 4. In 2035, public infrastructure projects will have been able to issue tokens to raise funds and engage the public in a regulated manner in some countries.

Projection 5. In 2035, tokenization-based solutions will have facilitated the establishment of decentralized and rule-based systems in governing and regulating infrastructure projects (e.g., permitting).

### (b) Economic projections

The potential economic efficiency gains brought by tokenization in infrastructure investment and development are some of the major factors to incentivize the industry to transition to a tokenized financing system. Even though these solutions have shown promises, many of them might not be eventually realized in practice. In order to investigate the economic benefits of tokenization can be realized in the short or long term and the prerequisites to enable adoption, seven projections are included in this section:

Projection 6. In 2035, tokenization will have been recognized as a viable option to be applied in both private and public infrastructure projects, which supplements existing infrastructure financing instruments.

Projection 7. In 2035, tokenization-based solutions will have been more frequently applied in public infrastructure projects than private projects.

Projection 8. In 2035, the cost of finance will have been reduced in infrastructure projects applying tokenization due to automation and dis-intermediation enabled by smart contracts.

Projection 9. In 2035, tokenization-based solutions will have contributed to the facilitation of cross-border transactions (either in the form of money, rights, or data), enabling international investors to invest in global infrastructure projects.

Projection 10. In 2035, institutional investors will prefer tokenization-based solutions to conventional instruments in financing infrastructure due to higher efficiency.

Projection 11. In 2035, the blockchain-based digital currency will have become dominant in payments in infrastructure projects applying tokenization.

Projection 12. In 2035, the liquidity of infrastructure investment will have been improved as tokenization-based solutions facilitate fractional ownership and unlock the marketplace for secondary trading.

### (c) Social projections

As discussed by researchers and practitioners, two of the key motivations for tokenization are that it allows for capital raise both from the global capital markets and from retail investors (including community residents) to participate in infrastructure development and that it leaves governments more accountable and transparent of investments in public assets and services. This shift could substantially enhance sustainability and inclusiveness, and six projections are included in this section:

Projection 13. In 2035, tokenization-enabled micro-transactions will have become standard due to reduced costs and efficient transaction processing, so that individuals (including surrounding community residents) can directly invest in infrastructure at the project level.

Projection 14. In 2035, infrastructure project stakeholders, regulators, or surrounding community residents will have been empowered in decision-making as tokenization makes some real-time project information (e.g., finance) more accessible.

Projection 15. In 2035, small and community-level infrastructure projects will have been able to raise funds directly from capital markets through tokenization.

Projection 16. In 2035, tokens embedded with certain rights (e.g., voting, access to services, financial return) distributed to impacted community residents will allow and encourage them to actively participate in local infrastructure development, thereby improving inclusivity and approval of new projects.

Projection 17. In 2035, tokenization-based solutions will have been more frequently applied in infrastructure projects in emerging countries with underdeveloped capital markets than in developed countries.

Projection 18. In 2035, tokenization-based solutions will have been applied to monetize positive social and environmental impacts, enabling new business models to promote sustainability and facilitate Environmental, Social, and Governance (ESG) investing.

### (d) Technological projections

Tokenization, as an emerging technology, is facing multiple technological barriers, which hurdle its broader adoption. Reliable technological support is of great importance to enable the realization of efficiency gains promised by the tokenization innovation. To explore the current technological status and the potential integration with other technologies, five projections are included in this section:

Projection 19. In 2035, transparency in infrastructure projects applying tokenization will have been improved as some project-related data are immutably stored on the blockchain and can be assessed in real-time.

Projection 20. In 2035, smart contracts applied in the tokenization of infrastructure will have been standardized.

Projection 21. In 2035, the tokenization of infrastructure will have been more frequently deployed on permissioned blockchain than on permissionless blockchain.

Projection 22. In 2035, tokenization-based solutions will have been in use in combination with other technologies (e.g., Internet of Things), enabling the establishment of an interconnected and intelligent infrastructure financing and managing system.

Projection 23. In 2035, the adoption of the tokenization of infrastructure will have reached sufficient scale, leading to the full realization of efficiency gains.

3.2. Selection of experts

An appropriate selection of experts avoiding bias is of paramount importance in Delphi studies as is a distribution of expertise and sufficient panel size to ensure reliability and validity. Since a heterogeneous expert base reduces cognitive biases and fits the interdisciplinary nature of this research topic, a great emphasis was placed on an adequate level of heterogeneity (Kopyto et al., 2020; Meldander et al., 2019). Heterogeneity was achieved through recruiting experts from various domains such as decentralized finance, asset tokenization, blockchain, and infrastructure finance, who were sourced from academia, industry, and political associations.

The selection criteria were publications in the domain, technical specialization in the related field, and expressed interest in the topics of infrastructure and tokenization. Experts selected to join the panel of this research include:

- Academics who have published peer-reviewed articles in the field of the application of blockchain in infrastructure investment and development
- Manager and executive-level experts in the infrastructure sector who have experience or been informed of tokenization.
- Managers and executive-level experts in the blockchain tokenization sector who have experience in infrastructure investment or have been informed of infrastructure as an asset class.

Experts were identified through personal networks, academic databases, and social network platforms (e.g., LinkedIn). Tokenization startups, Initial Coin offering (ICO), and Security Token Offering (STO) projects in the field of infrastructure development (e.g., energy sector) were also screened. The experts' personal recommendations were key sources in the search process. Academics were selected based on related academic publications and conference lists. Professionals were selected by considering individuals with above manager-level responsibility.

A scoring system was developed to reflect the individual expertise of experts (panelist candidates). Scores were based on a set of criteria, including management level, academic background, job specialization, education, publications, and years of experience. Experts with the highest score were invited to participate in this study. Consequently, 394 potential experts with expertise matched with the research topic were invited to join the panel (Day and Bobeva 2005; Nowack et al., 2011). Previous research shows significant differences in the number of experts included in a Delphi study, ranging from about twenty (Bokrantz et al., 2017) to several hundred (Fundin et al., 2018). As a result, no specific sample size target needs to be set (De Loë et al., 2016; Loo, 2002), as differences result from the specific topic and aim of each study (Culot et al., 2020; Kluge et al., 2020). In line with the cross-disciplinary and explorative nature of this study, panel size of 30-40 experts with heterogeneous backgrounds (a minimum of 10 for each subpanel) was chosen for this study based on general recommendations in Delphi research (Parente and Anderson-Parente, 1987; Borkrantz et al., 2017; Parente and Anderson-Parente, 2011). The response rate of 10.4% was found to be adequate for the research aim and similar to comparable Delphi studies (Spickermann et al., 2014; Warth et al., 2013). The final panel consisted of 39 experts who participated in the first Delphi round. Thirty-seven experts participated in the second round. This dropout rate of 5.1% compares well with other Delphi studies (Evans, 1997; Landeta, 2006; Landeta et al., 2008).

3.3. Execution of Delphi

This Delphi study is divided into several procedural steps. First, research panelists were asked to anonymously evaluate projections according to the probability and impact of occurrence.

Qualitative justifications to enrich quantitative assessments were used to amplify the interpretations of results. Subsequently, the panelists received feedback on the group's opinion and were offered the option to adjust their initial answers in the second round. Cluster analysis was used to develop scenarios, based on statistical analyses of responses and the qualitative arguments from the panel. The structure of the Delphi process is illustrated in Figure 1.

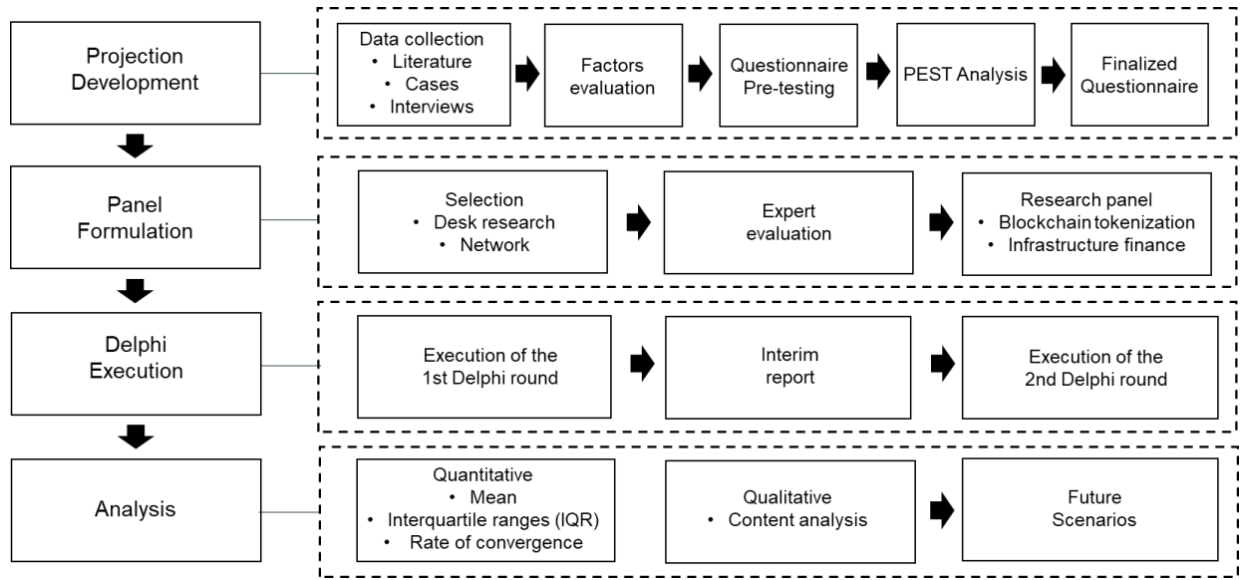

Figure 1. The structure of the Delphi process

This Delphi analysis was designed with two rounds. The purpose is to mitigate fatigue and ensure a high response rate. Numerous studies (Woudenberg, 1991) have found that three rounds or more might not increase the quality of the findings, but also increased the risk of panelist dropout (Fritschy and Spinler, 2019; Kluge et al., 2020). There is no optimal number of survey rounds in Delphi research, however, in the Delphi-based scenario studies, two rounds were found most commonly applied (Höjer, 1998; Mason and Alamdari, 2007; Piecyk and McKinnon, 2010; Melander et al., 2019).

Each Delphi round followed the same procedure. Experts were asked to assess the projections in accordance with

- Expected probability of occurrence (EP), based on a metric scale of 0–100%
- Impact on the industry (I), based on a 5-point Likert scale

Panelists were sent an email containing a unique link to the questionnaire through Qualtrics. Experts' responses could not be linked to their identities by individuals outside of the core research group to ensure privacy and confidentiality to ensure anonymity (Gill et al. 2013). For Round 1, the questionnaire was sent to the experts at the end of December 2021. The last questionnaire was returned in mid-February 2022 after three reminders. Before Round 2 evaluations, an interim report containing the algebraic mean of results and qualitative inputs from the Round 1 were sent to experts. After that, specific links to the questionnaire with individual results of the Round 1 were sent to each expert at the end of February 2022. Experts were allowed to modify their responses by opening the link. The last result received was in mid-March after three reminders. The aim of the second Delphi round was to increase consensus among experts by considering the other experts' feedback and to achieve higher data validity (Jiang et al., 2017; Heiko and Darkow, 2010). Three indicators suggest a high level of involvement and commitment in this study: (1) the entire research panel answered the entire questionnaire without leaving one projection blank; (2) a high number of, 446, usable comments were submitted; (3) the dropout rate (5.1%) is low, compared to the 18% drop-out rate (Nowack et al., 2011).

3.4. Analysis

To analyze and visualize the research panel's assessments regarding the expected probability and impact of the projections proposed, arithmetic mean values, interquartile ranges (IQR), and standard deviation (SD) were calculated. IQR indicates the consensus and dissent levels with 25 as the threshold level following the research of Warth et al. (2013) and Keller and von der Gracht (2014). IQR is measured only for the probability of occurrence. Convergence rates, measuring the difference in SD between Delphi rounds, show the panel's shift in assessments. Besides quantitative assessments, 672 written comments equivalent to 17.2 arguments per expert, are given by the panel. Seventy-four percent of the panelists provided at least one rationale to support their quantitative assessments. For the analysis of qualitative inputs, two researchers in the team worked as coders to structure the data and label arguments (Roßmann et al., 2018). Descriptive codes for the arguments were derived by applying the content analysis. Codes were discussed by two coders to readjust divergent coding results until consensus was achieved.

As suggested by (Saritas and Oner, 2004; Heiko and Darkow, 2010), the Delphi results were analyzed to derive scenarios according to quantitative results and qualitative arguments for further illustration and verification. A cluster analysis was conducted to analyze structures in the results. The variables considered in the cluster analyses were projections' mean values of probability of occurrence and impact. Numerous studies have argued that clustering based on these two dimensions is appropriate to derive appropriate strategies and actions (Ogden et al., 2005; Akkermans et al., 2003; Rikkonen et al., 2006). The results of the cluster analysis served as a basis to elaborate on scenarios for the application of tokenization in infrastructure investment and development in 2035. Further desk research was conducted to support the plausibility and consistency of the scenarios.

## 4. Discussions

4.1. Descriptive statistics

The descriptive statistics of the Delphi data are presented in Table 2. Experts' estimates of probability for 23 projections are depicted with the respective measurements of mean, convergence, and consensus in two Delphi rounds. The impact of projections is measured in both rounds but only the second-round evaluation results were used to develop scenarios for further analysis. The results revealed that the occurrence of all reviewed projections would have a non-negligible impact on infrastructure investment and development, as the mean values for the estimated impact were all rated greater than 3.0 on a 5-point Likert scale (Kopyto et al., 2020).

Thus, all projections were further considered for analysis. Projection 9 (cross-border transactions) and Projection 22 (Internet of Things integration) are anticipated to have the highest influence.

Table 2. Descriptive statistics analysis results

| No. | Projection | Probability Round 1 (N=39) | | | Probability Round 2 (N=37) | | | Round 2 vs. Round 1 | | Impact |
|---|---|---|---|---|---|---|---|---|---|---|
| | | Mean | SD | IQR | Mean | SD | IQR | Mean Change | CV | Mean |
| Political/Regulatory | | | | | | | | | | |
| 1 | Regulatory framework--most countries | 72.95 | 22.53 | 30.00 | 73.11 | 19.27 | 20.00 | 0.16 | -3.26 | 4.19 |
| 2 | Regulatory framework--international | 50.13 | 27.54 | 45.00 | 50.54 | 23.91 | 40.00 | 0.41 | -3.63 | 3.70 |
| 3 | Regulatory framework--smart contract | 70.00 | 26.83 | 45.00 | 69.05 | 26.06 | 35.00 | -0.95 | -0.77 | 4.08 |
| 4 | Regulated offering for public infrastructure | 76.28 | 21.85 | 32.50 | 78.51 | 18.78 | 20.00 | 2.23 | -3.07 | 4.27 |
| 5 | Rule-based governing system | 56.15 | 25.71 | 40.00 | 54.05 | 24.86 | 40.00 | -2.10 | -0.86 | 3.62 |
| Economic | | | | | | | | | | |
| 6 | Viable in private and public projects | 76.92 | 20.51 | 22.50 | 76.89 | 18.80 | 20.00 | -0.03 | -1.71 | 4.22 |
| 7 | More frequently applied in public projects | 36.28 | 24.49 | 30.00 | 32.30 | 19.42 | 15.00 | -3.98 | -5.07 | 3.16 |
| 8 | Reduced cost of finance | 75.13 | 26.67 | 40.00 | 77.57 | 22.69 | 35.00 | 2.44 | -3.98 | 4.22 |
| 9 | Cross-border transactions | 74.87 | 23.49 | 25.00 | 78.78 | 19.73 | 15.00 | 3.91 | -3.76 | 4.32 |
| 10 | Insitutional investors | 59.36 | 27.72 | 42.50 | 59.86 | 25.26 | 40.00 | 0.51 | -2.46 | 3.65 |
| 11 | Blockchain-based digital currency | 68.33 | 33.74 | 67.50 | 69.46 | 29.32 | 45.00 | 1.13 | -4.43 | 3.95 |
| 12 | Investment liquidity | 74.49 | 21.94 | 32.50 | 77.16 | 18.43 | 20.00 | 2.67 | -3.51 | 4.27 |
| Social | | | | | | | | | | |
| 13 | Micro-transactions | 70.26 | 26.73 | 42.50 | 74.05 | 20.94 | 30.00 | 3.80 | -5.78 | 4.00 |
| 14 | Real-time information access | 71.92 | 28.02 | 35.00 | 71.89 | 26.49 | 40.00 | -0.03 | -1.53 | 4.00 |
| 15 | Small-scale projects | 75.13 | 24.88 | 25.00 | 79.05 | 21.66 | 15.00 | 3.93 | -3.22 | 4.08 |
| 16 | Community engagement | 68.21 | 27.01 | 40.00 | 70.68 | 22.05 | 35.00 | 2.47 | -4.96 | 4.08 |
| 17 | More frequently applied in emerging countries | 57.31 | 26.23 | 35.00 | 57.97 | 23.11 | 30.00 | 0.67 | -3.12 | 3.81 |
| 18 | ESG investing | 78.59 | 23.34 | 35.00 | 81.35 | 17.70 | 20.00 | 2.76 | -5.64 | 4.27 |
| Technological | | | | | | | | | | |
| 19 | Transparency | 81.92 | 18.09 | 20.00 | 85.14 | 14.84 | 15.00 | 3.21 | -3.25 | 4.27 |
| 20 | Standardized smart contracts | 60.90 | 27.93 | 50.00 | 60.95 | 25.98 | 40.00 | 0.05 | -1.95 | 3.95 |
| 21 | Permissioned/Persmissionless blockchain | 62.95 | 30.94 | 45.00 | 64.86 | 25.91 | 35.00 | 1.92 | -5.03 | 3.51 |
| 22 | IoT intergration | 82.05 | 21.20 | 30.00 | 83.78 | 19.59 | 20.00 | 1.73 | -1.61 | 4.32 |
| 23 | Sufficient scale | 66.03 | 25.37 | 42.50 | 67.30 | 24.05 | 40.00 | 1.27 | -1.32 | 4.19 |

The mean values of the projections' probability range from 32.3% for Projection 7 (more frequently applied in public projects) to 85.1% for Projection 19 (transparency). The research panel increased its estimates in the second-round assessment regarding the probability of occurrence for 19 projections and decreased in four. For thirteen projections, the mean value of the probability is over 70%. Besides Projection 19, the estimated probability of another two projections, Projection 18 (ESG investing) and Projection 22 (IoT integration) is over 80%. Nine projections fall into the range from 50% to 70%. Projection 7 (more frequently applied in public projects) is the only projection below 50%.

Calculating the IQR of the research panel's probability assessments revealed that by applying the common threshold value of less than 25 as discussed in the Methodology section, Projection 6 (viable in private and public projects) and Projection 15 (small-scale projects) yield consensus in the first round. In the second round, ten projections yield consensus, which accounts for 43.5% of all projections. Besides the two aforementioned projections, Projection 1 (regulatory framework--most countries), Projection 4 (regulated offering for public infrastructure), Projection 7 (more frequently applied in public projects), Projection 9 (cross-border transactions), Projection 12 (investment liquidity), Projection 18 (ESG investing), Projection 19 (transparency), Projection 22 (IoT integration) yielded consensus in the second Delphi round. Nine out of 11 projections with the highest probability rate yield consensus. Projection 7 with the lowest probability rate also yielded consensus.

Experts' opinions converged over time as the convergence rates of all projections are negative. On average, the SD decreased by 3.2%, which indicates a convergence toward group consensus. The strongest convergence is observed in Projection 13 (micro-transactions), with a 5.8% decrease in SD. The lowest convergence rate, 0.77%, was measured in Projection 3 (Regulatory

framework--smart contract), which indicates that the participating experts were the most confident with the assessment of this proposition.

In addition to analysis on an aggregated panel level, further insights are given by a deep dive into subgroups' assessments, as illustrated in Table 3. The subgroups are classified by panelists' professional backgrounds and expertise (Roßmann et al., 2018). Experts in blockchain tokenization (n = 21) and infrastructure development or investment (n = 18) participated in this research. Infrastructure cohort assessments of the probability rates of 21 projections (91.3%) are lowered than those of the tokenization cohort. The idea behind comparisons between subgroups is that experts have unique perspectives and expectations based on their backgrounds (Kopyto et al., 2020). Projection 7 (more frequently applied in public projects) and Projection 22 (IoT integration) are the only two projections that the infrastructure cohort feels more optimistic about. Projection 19 (transparency) has the highest probability rated by the tokenization cohort while Projection 22 (IoT integration) is rated the highest by the infrastructure cohort. The infrastructure cohort is also more pessimistic about the impact of the projections. Projection 15 (small-scale projects) is the only one rated higher by the infrastructure group.

Table 3. Subgroup statistics analysis results

| No. | Projection | Proability Tokenization | Infrastructure | Impact Tokenization | Infrastructure |
|---|---|---|---|---|---|
| Political/Refulatory | | | | | |
| 1 | Regulatory framework--most countries | 78.81 | 65.63 | 4.19 | 4.19 |
| 2 | Regulatory framework--international | 56.90 | 42.19 | 3.90 | 3.44 |
| 3 | Regulatory framework--smart contract | 80.71 | 53.75 | 4.43 | 3.63 |
| 4 | Regulated offering for public infrastructure | 83.81 | 71.56 | 4.29 | 4.25 |
| 5 | Rule-based governing system | 61.90 | 43.75 | 3.81 | 3.38 |
| Economic | | | | | |
| 6 | Viable in private and public projects | 79.29 | 73.75 | 4.33 | 4.06 |
| 7 | More frequently applied in public projects | 31.19 | 33.75 | 3.19 | 3.13 |
| 8 | Reduced cost of finance | 85.48 | 67.19 | 4.33 | 4.06 |
| 9 | Cross-border transactions | 82.86 | 73.44 | 4.48 | 4.13 |
| 10 | Insitutional investors | 69.29 | 47.50 | 3.76 | 3.50 |
| 11 | Blockchain-based digital currency | 74.29 | 63.13 | 4.24 | 3.56 |
| 12 | Investment liquidity | 77.38 | 76.88 | 4.33 | 4.19 |
| Social | | | | | |
| 13 | Micro-transactions | 83.81 | 61.25 | 4.29 | 3.63 |
| 14 | Real-time information access | 77.38 | 64.69 | 3.95 | 4.06 |
| 15 | Small-scale projects | 82.14 | 75.00 | 4.05 | 4.13 |
| 16 | Community engagement | 77.14 | 62.19 | 4.14 | 4.00 |
| 17 | More frequently applied in emerging countries | 60.48 | 54.69 | 4.05 | 3.50 |
| 18 | ESG investing | 83.10 | 79.06 | 4.43 | 4.06 |
| Technological | | | | | |
| 19 | Transparency | 87.38 | 82.19 | 4.43 | 4.06 |
| 20 | Standardized smart contracts | 66.67 | 53.44 | 4.05 | 3.81 |
| 21 | Permissioned/Persmissionless blockchain | 67.86 | 60.94 | 3.62 | 3.38 |
| 22 | IoT intergration | 82.38 | 85.63 | 4.33 | 4.31 |
| 23 | Sufficinet scale | 74.52 | 57.81 | 4.38 | 3.94 |

4.2. Scenario analysis

Scenarios of probable future, scenarios of possible future, and scenarios of surprising future were analyzed and are shown in figure 2. The figure maps the probability of occurrence against the impact of the projections are likely to have. Experts' comments on each projection are summarized in Appendix A.

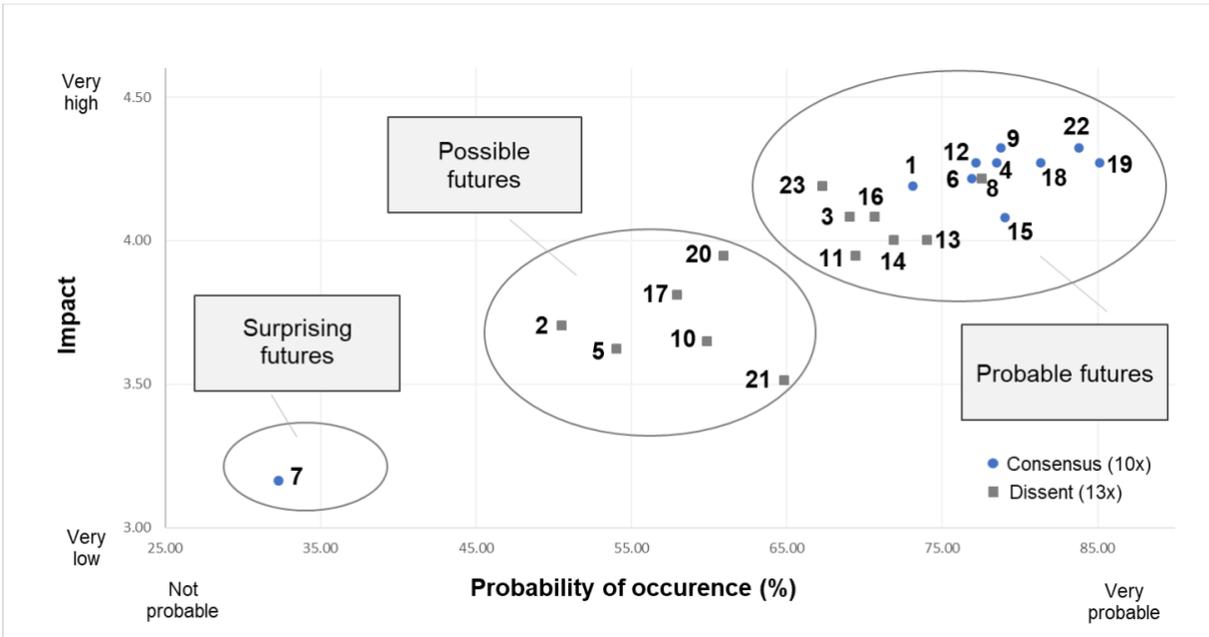

Figure 2. Scenarios of probable, possible, and surprising futures

4.2.1. Scenarios of the probable future

The scenarios of the probable future contain 16 projections that are characterized by the highest expected impact and probability of occurrence assessed by experts. The probability ranges between 67% (Projection 23) and 85% (Projection 19). The impact ranges between 3.95 (Projection 11) and 4.32 (Projection 22). Eight out of nine projections with the highest probability rate yield consensus. Projection 8 is the only exception. The high probability, impact, and consensus rate indicate convergence in these scenarios for 2035.

*Efficiency gains:* Experts are very optimistic that tokenization will improve transparency in infrastructure projects, as it is the highest-probability projection with a very high impact rate (Projection 19). Transparency is one of the main characteristics of blockchain-based solutions (Akkermans et al., 2021), which also empowers project stakeholders, regulators, and community residents in decision-making due to the real-time information disclosure capability (Projection 14). Data recorded on public blockchains are immutable and open to anyone with internet access. The cost of finance reduction in infrastructure projects applying tokenization due to automation and dis-intermediation enabled by smart contracts is also discussed by experts as a very likely efficiency gain (Projection 8). However, material advancements in smart contract design and enforcement are prerequisites to realizing cost reductions. The research panel predicted that it is likely that blockchain-based digital currency will have become dominant in payments in infrastructure projects applying tokenization by 2035 (Projection 11). Central bank digital currency (CBDC) issued by governments and other forms of digital currency based on blockchain which can be seamlessly integrated with the tokenized system are expected to be the norm by then. According to the International Monetary Fund (IMF), nearly 100 countries are exploring CBDCs as of early 2022, but only 9 countries have fully implemented them (Georgieva, 2022). However, some experts expressed concerns that conventional banking and government systems will not let loose control of the payment system easily. Liquidity improvement is one of the most frequently discussed benefits associated with tokenization. The research panel predicted with a high level of confidence (high probability and consensus reached) that tokenization-based solutions will contribute to the facilitation of cross-border transactions (enabling investors worldwide to invest in global infrastructure projects) (Projection 9) and that the liquidity of infrastructure investments will have been improved as tokenization-based solutions facilitate fractional ownership and unlock marketplaces for secondary trading (Projection 12). Potential investors in infrastructure investment can be expanded from institutional investors only to include retail and SME investors (domestic and international) through tokenization. The decentralized finance (DeFi) momentum in the marketplace will also keep attracting more investors to participate in tokenized investments. The potential integration of tokenization with other technologies is rated positively by the research panel. The projection that tokenization-based solutions will have been in use in combination with other technologies (e.g., Internet of Things), enabling the establishment of interconnected and intelligent infrastructure financing and management system is rated the second highest possibility and yielded consensus. The combination of IoT and blockchain has already been underway. Numerous interesting use cases in other areas (e.g., supply chain management) are currently being explored. Data generated from the IoT devices can be stored on the blockchain directly and automatically through oracles (Woo et al., 2020).

*Regulatory environment:* Proper regulation is essential to the safe development and implementation of the tokenization of infrastructure, which promotes stability, protects market participants, and ensures integrity. The research panel has a high level of confidence that legal and regulatory frameworks for the tokenization of infrastructure will have been established in most (developed and developing) countries by 2035 (Projection 1). Some countries have established legal regulations for tokenized securities (e.g., Liechtenstein and Switzerland) and it is an ongoing trend (Culinovic-Herc et al., 2021). Experts believe that tokenization merely replaces one digital technology with another, which should not raise issues in jurisdictions to have it regulated under the existing framework. However, the high level of sophistication of infrastructure projects can be an obstacle for tokenization implementation, and developing countries are expected to lag behind in regulating this sector. The clarification of smart contracts in legal and regulatory frameworks is supported by the research panel (Projection 3), as smart contracts fit into the existing framework of the contract and regulatory law regimes. The research panel supported and achieved consensus that public infrastructure projects will have been able to issue tokens to raise funds and engage the public in a regulated manner in some countries in 2035 (Projection 4). Enabling public projects to raise funds through a token sale should achieve similar goals of issuing municipal bonds under the existing public financing system and legal framework.

*Social impacts:* Besides economic impacts made by tokenization-based solutions, potential social impacts also draw public attention. The research panel predicted that individuals (including community residents) can directly invest in infrastructure at the project level through micro-transaction enabled by tokenization by 2035. The new investment opportunities will attract retail investors to add infrastructure investment into their portfolios and encourage community residents to actively participate in local project development. However, if the acquisition of tokens is only by means of capital, which could assign more rights to the rich. An inappropriate design might actually further increase the wealth gap and worsen inclusivity. Some experts argued that even if the mechanism allows retail investors to participate, they (including community residents) do not really have the characteristics for such investments. So, the anticipated rate of participation can still be low if it can happen in reality. The research panel reached a consensus and rated the high possibility for small and community-level infrastructure projects will have been able to raise funds directly from capital markets through tokenization in 2035 (Projection 15). The disintermediation enabled by smart contracts results in a significant fixed financing cost reduction and improved bankability of projects. The scale of projects or corporations is no longer a determining factor for access to financing in public markets. Experts were optimistic that tokenization would make capital market fundraising for infrastructure projects at the community level possible. Some experts expressed the need for a regulatory framework to enable and facilitate the realization of community-level financing. Engaging surrounding residents in the impacted community through tokenization was also rated likely by the research panel (Projection 16). Community-level Decentralized Autonomous Organizations (DAO) can facilitate the implementation. Experts were excited (high probability, high impact, and consensus yielded) about applying tokenization to monetize positive social and environmental impacts, enabling new business models to promote sustainability and facilitate Environmental, Social, and Governance (ESG) investing (Projection 18). The urgent need for implementing emerging technology to facilitate the integration of ESG metrics in investments may result in faster adoption in relevant fields. Tokenization could allow fundamental transparency to be built into ESG initiatives, which opens the door for scaling monetization and incentivization techniques.

*Technology adoption:* Experts were asked to evaluate the general adoption of tokenization in infrastructure projects. The research panel leaned to support the adoption of tokenization of infrastructure to reach sufficient scale, leading to the full realization of efficiency gains by 2035 (Projection 23). However, the probability rate (67.3%) is the lowest among projections in the group of scenarios of the probable future. The consensus was not achieved for this projection, which indicated a relatively lower level of confidence in this group. Sufficient scale is hard to define, besides it is highly uncertain due to regulatory technological obstacles whether tokenization could reach the "tipping point" of mass adoption to fully realize efficiency gains. Some types of infrastructure projects tend to be more suitable for tokenization (utility infrastructure) and reaching sufficient scale in these projects would be more likely. Experts predicted that tokenization would have been recognized as a viable option to be applied in both private and public projects, which supplement existing infrastructure financing instruments by 2035 (Projection 6). There have been some use cases of tokenization in private projects (e.g., energy infrastructure). Some experts believed that once tokenization is successfully demonstrated in some pilot public projects, other entities are expected to quickly follow, pending regulatory approval.

4.2.2. Scenarios of the possible future

The scenarios of the possible future contain six projections characterized by the medium expected impact and probability of occurrence assessed by the research panel. It is important to incorporate eventualities that have a medium expected probability of occurrence and a rather high impact as they indicate directions of the alternative future (Grossmann et al., 2007). The probability ranges between 50% (Projection 2) and 65% (Projection 21). The impact ranges from 3.51 (Projection 21) to 3.95 (Projection 20). Six projections in these scenarios are characterized by high IQR, indicating a higher degree of disagreement among experts and no consensus was reached. Even though the experts were not in agreement and their assessments do not show any tendency toward a low or high probability and impact, their comments reveal essential points that may influence the future landscape of blockchain-enabled tokenization utilization in infrastructure investment and development.

*Global adoption:* Enabling the adoption of tokenization at the global level could expand the investor base and link projects worldwide with global markets. An international regulatory framework for the tokenization of infrastructure is essential and the prerequisite to realize efficiency gains (Projection 2). However, experts had different opinions on whether an international regulatory framework could be established. Countries have distinct regulatory environments, limiting the opportunity to harmonize different parties' interests at a global level in spite of the great introduced impact. Some experts were optimistic that regional or bi- and multi-lateral regulatory frameworks (e.g., within the EU) are more likely to be established. The gap between developing and developed countries is another obstacle faced by the global adoption of tokenization (Projection 17). Indeed, developing countries have fewer burdens so they are more flexible to find alternative paths to finance their infrastructure projects (Tian et al., 2021), while developed countries are vested in the current financing mechanism. Whether this is due to governments and institutional centralized fear of the loss of power, as commented by some experts, or barriers to change in processes, is unclear. However, many developing countries may not have the required economic growth, and stable regulatory and financial systems to enable the transition. Experts tend to believe that the tokenization of infrastructure will be more likely to be applied in developed countries with limited pioneering by developing countries.

*Decentralized governance:* The process of permitting under the current centralized system with human intervention creates a series of obstacles to infrastructure development, such as briberies, holiday delays, and political influences. The decentralization enabled by blockchain solutions might enable the creation of automatic algorithm-based governance and permitting system. As long as pre-defined conditions in smart contracts are met, permits can be automatically granted to relevant parties. Overall, experts expressed divided opinions on Projection 5. The supporting experts argued that automation could be achieved in some parts of the permitting or regulatory processes and a DAO could facilitate the transition. Others stated that it is rather challenging to remove authority from governments, and legacy systems will not allow relinquishment of control. Permitting and regulating is rather complicated, it would be unlikely to have a system developed at least in the short term to replace human-based decision making.

*Institutional investor:* Institutional investors are key players in infrastructure investment and development under the current financing models. The research panel was not very optimistic that institutional investors would prefer tokenization-based solutions to conventional instruments in financing infrastructure even at purported higher efficiency (Projection 10). Tokenization could expand the investor base to retail and SME investors, who are considered competitors in infrastructure investment. While some experts predicted if efficiency gains brought by tokenization were equal to or greater than conventional instruments at equivalent risk, tokenization-based solutions may be used to complement other financial instruments by institutional investors.

*Technical support:* The discussion on whether future tokenization activities will be deployed on permissioned or permissionless blockchain reflects further disagreement (Projection 21). Some experts predicted that permissioned blockchain would be more frequently applied initially due to better control and regulated markets might only be permissioned under the current legal system. While it is argued that permissionless blockchains have a broad appeal to a larger base of users and the impact of the deployment of permissionless blockchains to enable a decentralized system is rather high. It is unclear which blockchain will be more frequently deployed based on the assessments of this research. Whether smart contracts designed for the tokenization of infrastructure would have been standardized in 2035 has a great impact on efficiency realization (Projection 20). The research panel predicted that it might be challenging to have standardized smart contracts designed for the entire infrastructure sector due to the complexity and variety of projects. Partial standardization for certain sectors is more feasible.

### 4.2.3. Scenarios of the surprising future

The scenarios of the surprising future contain only one projection characterized by the rather low expected impact and probability of occurrence assessed by the experts. Projection 7 is the only projection in this scenario. The probability of Projection 7 is the lowest among all projections, which is 32%. Its impact, 3.16, is also the lowest. Consensus was achieved in this projection. This scenario describes surprising (unlikely) developments associated with tokenization in infrastructure investment and development.

*Application in public or private projects:* The low possibility rate of Projection 7 indicates that tokenization-based solutions will have been more frequently applied in public infrastructure projects than private projects by 2035 is not supported by the research panel's assessment. Experts predicted that tokenization would be more likely to be applied to the private sector in less than 15 years of time. The consensus shows a high level of confidence in experts assessments. The lowest impact among all projections indicates that whether the project is public or private is not vitally important to the tokenization application. As commented by experts, the public sector always lagged in adopting emerging technologies and innovations due to the top-down decision-making and bureaucratic management systems. A stricter regulatory environment faced by public projects and the low-risk appetite (security sometimes is the priority instead of efficiency) is also the reason why tokenization is more likely to be implemented in the private sector at the early stage of adoption. Nevertheless, experts also predicted that infrastructure privatization and innovative PPP projects in the public sector have a better chance for tokenization as compared to other types of public projects. It is not negligible that the majority of infrastructure projects are in still public hands, even though a small portion of these projects apply tokenization, the overall number of public projects applying tokenization can be still higher than the application in private projects. It was also mentioned by some experts that public infrastructure assets have appeal to many investors, but are not available for most of them. The demand for investing in public infrastructure through tokenization under innovative PPP contracts could facilitate broader adoption. Once the security and efficiency of tokenization are proven by its application in the private sector and with the support of regulation, its adoption in the public sector can be rather quick, despite the fact it takes a longer time.

### 4.3. Implications for policy, practice, and research

Proper regulations and policies are essential to the future applications of tokenization in infrastructure investment and development. They are prerequisites to further adoption and realizations of efficiency gains. Regulatory and supervisory authorities should take actions to regulate the tokenized market to prevent fraudulent activity within the ecosystem, promote market integrity, safeguard against misconduct risks, and provide clear guidance to allow participants to innovate in the market. For policymakers, the results of this research indicate the regulatory and policy framework for the tokenized market is still in its infancy. Regulations specific to the tokenization of infrastructure have not been established. The legal status of digital (crypto) tokens and smart contracts should be recognized. Tax legislation should also be modernized to take on changes resulting from the tokenized investment. Allowing the public (including surrounding residents) in decision-making and public infrastructure projects to raise capital from the markets directly through tokenization would benefit the overall society. If developing countries can seize the opportunity to regulate the tokenization market sooner, considering fewer systematic burdens faced, they are more flexible to find an unusual way to finance their infrastructure projects and elevate their economies. Educational efforts are vital to the border adoption of tokenization, especially for the public to understand the emerging technology, which should be addressed by the policymakers. There is a continuous need for updating the existing legal and regulatory frameworks to address the changes brought by tokenization-related activities. While policymakers should also ensure that regulations are not too stringent to hinder innovation.

For practitioners, this study offers a validated starting point in a field characterized by rather uncertain conditions and serves as an inspiration for future developments. The tokenization of infrastructure has been explored in some private projects mainly in the energy sector. As experts predicted it would be not only possible but also make high impacts if tokenization can be applied in public projects, which should be proactively investigated as the next step of broadening tokenization in infrastructure to better serve communities and the general society. The reduced cost of finance relies on sufficient technological support (well-developed and standardized smart contract), which can only be realized once the scale of adoption reaches a certain level. Even though the investor base of infrastructure projects is enlarged to include international and retail investors in theory, risk and return should be properly designed and structured to engage these investors to participate in practice. Tokenization could bring both benefits and challenges to institutional investors. It is important for them to get familiar with emerging technology through pilot projects. Liquidity improvement is also considered a major benefit associated with tokenization. The establishment of regulated token markets and well-implemented investor protections are prerequisites. Enabling decision-making for the public in infrastructure through corporate services (e.g., voting) is considered possible by applying tokenization. However, what kind of access should be granted to balance the benefits and obstacles should be well examined. It is necessary for practitioners to allocate more efforts and resources to investigate the integration of tokenization with other technologies to promote ESG investment to address the urgent demand at present.

Implications from the perspective of the methodology include the contribution to the development of forecasting methods by applying the Delphi technique. Although Delphi and scenario planning have been applied in many studies, it has yet to be used in the field of blockchain-enabled tokenization. This study provides evidence of the importance of Delphi as a forecasting toolforo technology adoption studies. Considering tokenization is in its infancy, there is limited data available for in-depth analysis at present. Future research should conduct analysis based on real-time data (from operations) to explicitly examine to what extent and how tokenization can realize the projected efficiencies gains (comparing tokenized finance with conventional finance options). The research on how tokenization can be integrated with project finance would contribute to the infrastructure finance literature significantly. As discussed by some researchers, the Delphi analysis is normally part of a more comprehensive research process (Rowe & Wright, 2011). While results of the Delphi analysis provide insights into possible future developments, they might not be generalizable or enable theory-building. Further research should explore results in more depth and balance out the limitations of the future-oriented research methods if possible. The research panel consists of 39 experts as experts who have experience in this emerging field are very limited. Future research can place more emphasis to conduct a more in-depth analysis after there are more practitioners and scholars working on the tokenization of infrastructure. It is also important to distinguish the application of tokenization in different types of infrastructure projects (e.g., transportation, energy), which might yield interesting additional insights. The use of additional forecasting techniques catering to short-term predictions and qualitative empirical work that monitor the developments caused by blockchain in society and markets could enrich the findings of this study.

## 5. Conclusions

Spurred by the emerging blockchain technology and increased interest in tokenization, this forecasting research is built on extended literature and aggregated expertise to investigate the potential implementation of blockchain-enabled tokenization in infrastructure investment and development. While blockchain technology forecasting for financial services has been conducted (Chang et al., 2020), the research on the field of infrastructure is highly theoretical and in its very early stages (Wamba and Queiroz, 2020), there is an urgent need for further empirical analysis to better understand efficiency gains, implementation processes, challenges, and future probable application areas. The Delphi-based scenario analysis approach was applied to gather long-term forecasts and assessments of a research panel consisting of 39 experts in blockchain tokenization and infrastructure development on how tokenization will influence the future of infrastructure finance and identify future scenarios of potential applications in this field. Twenty-three projections developed from a literature review, case study, and experts' interview, concerning perspectives of both the supply and demand side, for the year 2035 were assessed in a two-round Delphi analysis. Regulatory, economic, social, and technological perspectives of tokenization were taken into consideration. Three groups of scenarios were created based on quantitative (the probability of occurrence and impact of occurrence rated by the research panel) and qualitative analysis (verbal inputs), which offer thought-provoking and holistic insights of tokenization's future in infrastructure. For organizations and investors that have had experience in the tokenization of infrastructure, the data of the Delphi analysis and scenarios can provide validation or expansion of the existing knowledge and system. For organizations and investors that have interests in the tokenization of infrastructure, the analysis and scenarios might be used to update or develop new strategies for future business plans. The results of this study clearly underlined the potential of tokenization. Uncertainties and barriers confronting the technologies' diffusion were discussed. This study contributes to the transfer of general technical-driven blockchain and tokenization knowledge to infrastructure-specific tokenization knowledge. It is essential since technology transition does not only depend on technological aspects alone but also depend on political, economic, and social aspects. The findings of this research provide policymakers, practitioners, and researchers with the first guideline to develop a blockchain-based strategy. By revealing the underlying value drivers, this study provided a starting point to identify relevant application fields for the technology. Long-term strategic planning is supported by this study with the scenario data acting as a starting point for long-term blockchain-related efforts in infrastructure development.

**Appendix A**

**Qualitative results**

P1. In 2035, legal and regulatory frameworks for the tokenization of infrastructure will have been established in most (developed and developing) countries.

High probability comments

- Some countries have established legal regulations for tokenized securities (e.g., Liechtenstein and Switzerland) and it is an ongoing trend.
- Tokenized solutions have proven their value to stay which leaves governments with no choice but to legislate.
- Tokenization merely replaces one digital technology with another, which should not raise issues in jurisdictions to have it regulated under the existing framework.
- The tokenization of infrastructure is included in a broader tokenization-based regulation.

Low probability comments

- It takes a longer time to develop legal and regulatory frameworks for infrastructure-specific tokenization while a general tokenization framework is more likely to be established before 2035.
- Most developing countries are likely lagging again in regulating tokenization even though some small countries might rush ahead in this matter.
- Legal frameworks take time and effort to develop and require a high level of sophistication.

P2. In 2035, an international regulatory framework for the tokenization of infrastructure will have been developed.

High probability comments

- Some initiatives related to the regulation of tokenization at the global level have been taken.
- A regional regulatory framework (e.g., within the EU) is more likely to be established.
- International organizations (e.g., World Bank and UN) can take the lead and implement it.
- A regulatory framework established by a few pioneering countries can be accepted by other countries.

Low probability comments

- Regulation related to tokenization is not easy to harmonize at a global level in spite of the fact that this would have a great impact on large-scale infrastructure projects.
- Countries have their own distinct visions, as a result, geopolitical disagreement and competition make it less likely for an established international framework.
- Governments have little incentives to collaborate to establish an international framework
- Not all countries (particularly big countries) agree with the international regulatory framework approach even in well-understood commercial areas,

P3. In 2035, legal obligations and regulations of smart contracts will have been clarified in most countries.

High probability comments

- It is highly likely that this will occur given the widespread use of smart contracts and this is high on the agenda of policymakers already today.

- The need for the smart contract is appreciated by the industry which is eager for efficiency, and the implementation is also more intuitive.

Low probability comments

- Smart contract functionalities are always evolving. Even if regulations around one type (e.g., Decentralized Finance) are resolved, something else will come up that requires new legal interpretation.
- Smart contracts and their applications require a level of legal sophistication, which takes time to have policymakers trained to understand the emerging technology before they can even start interpreting these obligations

P4. In 2035, public infrastructure projects will have been able to issue tokens to raise funds and engage the public in a regulated manner in some countries.

High probability comments

- The fundraising of public or private projects through tokenization is similar. If implemented successfully in public projects, the involvement of the public will change the industry fundamentally.
- Public entities are not able to resist this easy method of fundraising under the condition that a legal foundation has been created first.
- Raising through a token sale achieves similar goals of issuing municipal bonds under the existing financing system, which should be understood by upcoming generations of politicians

Low probability comments

- Infrastructure financing is still relatively complex in itself; the integration of tokenization and infrastructure finance could only happen in less than a handful of countries.
- The existing system resists changing; tokenization might be considered a threat instead of an opportunity.

P5. In 2035, tokenization-based solutions will have facilitated the establishment of decentralized and rule-based systems in governing and regulating infrastructure projects (e.g., permitting).

High probability comments

- The use of tokens for governance and administration-related issues is more likely than the use of tokens for financing.
- Tokenization and smart contract could help need to shorten the environmental and other regulatory review process to reduce project development duration and cost.
- It is more likely to be realized with the support of a Decentralized Autonomous Organization (DAO).
- The emerging technology is evolving to be implemented in more complicated situations.

Low probability comments

- Changing conditions make it harder to establish a rule-based system beforehand.
- It is rather challenging to take authority away from governments. Legacy entrenchment will not allow relinquishment of control.

P6. In 2035, tokenization will have been recognized as a viable option to be applied in both private and public infrastructure projects, which supplements existing infrastructure financing instruments.

High probability comments

- Some countries have implemented tokenization in both private and public projects. The trend is ongoing.
- Once tokenization is successfully implemented in pilot projects, other entities will follow.
- Tokenization serves as a low-cost alternative to some existing financial instruments.

Low probability comments

- Regulations in this field should have been clarified.
- It is more difficult to have tokenization implemented in public projects.

P7. In 2035, tokenization-based solutions will have been more frequently applied in public infrastructure projects than private projects.

High probability comments

- Investors' demand for public infrastructure assets can facilitate and expedite the realization
- Tokenization is likely to be implemented in public-private partnership projects
- As most infrastructure are public projects, it is expected that tokenization will be used there but the frequency of the implementation of tokenization in private projects is higher.
- Tokenization can be implemented in privatization projects

Low probability comments

- Public entities reluctant to accept innovations
- Public infrastructure requires a dismantling of bureaucracy, for this reason, the implementation of tokenization in the public sector will be behind compared to the private sector.

P8. In 2035, the cost of finance will have been reduced in infrastructure projects applying tokenization due to automation and dis-intermediation enabled by smart contracts.

High probability comments

- Cost reduction through automation and disintermediation enabled by smart contracts is considered a major efficiency gain brought by tokenization.

- The cost of finance may be mostly reduced by a better engagement of the public and fewer objections to the relevant project.

Low probability comments

- There will have to be material advancements in smart contracts to achieve those cost reductions.
- This will have to be supported by public sector infrastructure - particularly the legal recognition of smart contracts.

P9. In 2035, tokenization-based solutions will have contributed to the facilitation of cross-border transactions (either in the form of money, rights, or data), enabling international investors to invest in global infrastructure projects.

High probability comments

- Global investors have already been subscribing to crypto-based solutions.
- There have been some companies offering cross-border investment services based on tokenization.
- Government-based cryptocurrency (e.g., central bank digital currency) will play an important role in international transactions.

Low probability comments

- It will be more difficult from a legal standpoint to introduce tokenization to cross-border infrastructure projects
- Governments tend to take tight control over foreign investment.
- Anything that challenges the existing authority of government is difficult to be implemented.

P10. In 2035, institutional investors will prefer tokenization-based solutions to conventional instruments in financing infrastructure due to higher efficiency.

High probability comments

- As tokenization introduces new ways of asset management, it will be a standard model of financing accepted by investors.
- It happens when efficiency gains are equal to or greater than conventional instruments at an equivalent risk
- It unlocks new opportunities for institutional investors to grow their revenues.

Low probability comments

- Institutional investors might be reluctant to accept tokenization as it introduces other investors (e.g., retail) to compete for investment opportunities.
- The impediments at the current stage are still too high to have this projection realized before 2035.

- The syndication of financing creates additional issues for institutional investors to consider tokenization.
- Institutional investors will be slow to adopt the technology only under the condition that they have understood the security aspects of tokenization before they focus on higher efficiency.

P11. Blockchain-based digital currency will have become dominant in payments in infrastructure projects applying tokenization.

High probability comments

- Central bank digital currency and other blockchain-based digital currency (e.g., cryptocurrency, stablecoins) will be the norm by 2035.
- Dividend distributions and other corporate actions can be managed in the best way by using digital currencies.
- The means of payment for tokens facilitate the process but might not make much of a difference for big investors (as opposed to speculators).
- Better efficiency will lead to quicker adoption.
- Digital currency is seamlessly integrated with the tokenized system.

Low probability comments

- Blockchain-based digital currency will be more likely complementary but not dominant.
- The conventional banking and government system will not let loose control of the payment system easily.
- Blockchain-based digital currency doesn't have the capacity to handle high-volume transactions.

P12. In 2035, the liquidity of infrastructure investment will have been improved as tokenization-based solutions facilitate fractional ownership and unlock the marketplace for secondary trading.

High probability comments

- Utility infrastructure (e.g., energy, water) will benefit the most.
- International investors and markets are unlocked.
- Liquidity improvement is considered the most disruptive impact of tokenization.
- The Decentralized finance momentum will keep attracting investors to participate in the tokenized investment.

Low probability comments

- A massive adoption of tokenization for a large number of infrastructure projects to allow for a real marketplace to be created is a pre-condition.
- Tokenized infrastructure doesn't fit the appetite of the tokenization investors.

P13. In 2035, tokenization-enabled micro-transactions will have become standard due to reduced costs and efficient transaction processing, so that individuals (including surrounding community residents) can directly invest in infrastructure at the project level.

High probability comments

- Regulators will facilitate the process even for unaccredited investors, as this is a market where micro-transactions or micro-investments are possible.
- The new investment opportunities will attract retail investors to add infrastructure investment into their portfolios.

Low probability comments

- Retail investors do not really have the characteristics for such investments; the anticipated rate of participation should be low even if the mechanism allows them to participate.
- It's not easy to become the new standard in 15 years, even in this fast-moving era and sector.
- The projected reduced costs and efficient transaction processing will still not be enough to allow microtransactions to happen.
- Social investing will not be a driving force for tokenization.

P14. In 2035, infrastructure project stakeholders, regulators, or surrounding community residents will have been empowered in decision-making as tokenization makes some real-time project information (e.g., finance) more accessible.

High probability comments

- This is one of the quickest wins of tokenization related to increased transparency due to the characteristics of blockchain.
- Investors can participate in decision making through Decentralized Autonomous Organization (DAO) platforms

Low probability comments

- Allowing the community to participate will be heavily regulated to prevent decision-making from impacting project delivery.
- In spite of fact that real-time information will be more accessible, how to understand, interpret and act upon that additional data remain questionable.

P15. In 2035, small and community-level infrastructure projects will have been able to raise funds directly from capital markets through tokenization.

High probability comments

- It can only be realized through tokenization, which has a great impact on infrastructure.
- Utility infrastructure is more likely to implement tokenization.

Low probability comments

- Capital markets have never been a major financing source for infrastructure regardless of scale.
- It is up to the regulation.
- It is more likely bounded in a fund and less likely realized on a project level.

P16. In 2035, tokens embedded with certain rights (e.g., voting, access to services, financial return) distributed to impacted community residents will allow and encourage them to actively participate in local infrastructure development, thereby improving inclusivity and approval of new projects.

High probability comments

- It is rather easy to be realized through blockchain-based solutions, such as Decentralized applications.
- It is easier to be implemented in communities with proactive residents.
- Community-level Decentralized Autonomous Organizations can facilitate the implementation.

Low probability comments

- A mindset shift has to happen first
- The danger with any crowdfunded infrastructure is that wealthier areas have more projects funded which further increases the wealth gap, actually worsening inclusivity.
- The acquisition of tokens if by means of capital will assign the rights to the rich, which might not change the current market dynamics.
- Institutional investors will prevent this from happening.
- Retail investors don't have the expertise.

P17. In 2035, tokenization-based solutions will have been more frequently applied in infrastructure projects in emerging countries with underdeveloped capital markets than developed countries.

High probability comments

- It is more like to happen in emerging countries with democratic governments.
- Developing countries have fewer burdens so they are more flexible to find an unusual way to finance their infrastructure projects, while developed countries are struggling with the current financing mechanism since institutions fear the loss of power.

Low probability comments

- It will be more frequently applied in small-sized but developed countries such as Switzerland, Denmark, etc., or distinctive nations like Saudi Arabia.
- It is very hard to evaluate; A lot of activities in blockchain-based applications in finance have been witnessed happening in underdeveloped countries given they have the greatest need for such efficiencies, but the most important pilots are still being run in advanced economies.

- Sometimes emerging markets can surpass and adopt new technology, but most countries do not have a solid, stable, sound economic status to make the transition.
- The financial system always leaves undeveloped countries undeveloped.

P18. In 2035, tokenization-based solutions will have been applied to monetize positive social and environmental impacts, enabling new business models to promote sustainability and facilitate Environmental, Social, and Governance (ESG) investing.

High probability comments

- The urgent need will result in faster adoption.
- The interpretation of the internet of things can provide real-time data that can be monetized (e.g., carbon-traded) and used by asset managers to meet ESG targets or to attract thematic investors.
- ESG investing may require clear rules which need to be fulfilled. Smart contracts can help to make ESG investing more feasible.
- This is one of the areas I am most excited about. Tokenization will allow fundamental transparency to be built into ESG initiatives, which opens the door for countless monetization/incentivization techniques.

Low probability comments

- It is hard to be realized due to the complexity of ESG investing.
- It lacks support from the demand side.

P19. In 2035, transparency in infrastructure projects applying tokenization will have been improved as some project-related data are immutably stored on the blockchain and can be assessed in real-time.

High probability comments

- It is a main characteristic of blockchain-based solutions
- The improved transparency will shape the governance process.

Low probability comments

- Developers and regulators are sensitive to certain information disclosure.

P20. In 2035, smart contracts applied in the tokenization of infrastructure will have been standardized.

High probability comments

- It will be easily realized in some sectors of infrastructure while might not be feasible in other sectors.
- Partial standardization can be realized sooner.

- Once tokenization has been widely applied in the infrastructure, it won't take much time to have standardized smart contracts developed.

Low probability comments

- It is an ongoing process but remains a difficult objective due to the constantly changing environment.
- Flexibility to address differentiation will be more important than standardization.
- It has been already difficult to determine what the base network is to rely on for future activities there is no need to say how difficult it is to standardize smart contracts for various types of infrastructure.
- Project finance has not been standardized after years of effort.

P21. In 2035, the tokenization of infrastructure will have been more frequently deployed on permissioned blockchain than permissionless blockchain.

High probability comments

- Institutional investors will prefer permissioned blockchain as security is their priority while it is unclear which type of blockchain is preferred by retail investors.
- Permissioned blockchain will be applied more frequently initially due to better control.
- Regulated markets will only be permissioned.
- Companies don't want to lose control.

Low probability comments

- Permissionless blockchains have a broad appeal to a larger base of users
- Tokenization on a permissioned blockchain would basically negate the benefits of tokenization because the tokens would not be interoperable with public chains and therefore cannot tap any of the accompanying financial benefits.
- The impact of the deployment of permissionless blockchain to enable a decentralized system is very high.

P22. In 2035, tokenization-based solutions will have been in use in combination with other technologies (e.g., Internet of Things), enabling the establishment of an interconnected and intelligent infrastructure financing and managing system.

High probability comments

- The combination of IoT and blockchain is already on its way and numerous interesting use cases in other areas (e.g., supply chain management) are currently being explored.
- Blockchain and IoT are complementary technologies, it would be very surprising if they didn't evolve to supplement one another
- Data generated from the Internet of things devices can be stored on blockchain automatically through oracles.

Low probability comments

- The integration with the Internet of things represents the past but not the future.
- Human indispensable is necessary when unexpected issues occur.
- Regulatory obstacles need to be overcome.

P23. In 2035, the adoption of the tokenization of infrastructure will have reached sufficient scale, leading to the full realization of efficiency gains.

High probability comments

- Fifteen years are long enough to prove the best use cases with the most efficiency gains.
- Once the tipping point is reached, wide adoption can be realized in a rather fast speed.

Low probability comments

- It is difficult to determine what the sufficient scale actually means; efficiency gains can be gained for each individual project that applies it.
- It varies depending on the type of projects they are.